# Raman spectroscopy of graphene at high pressure: substrate and pressure transmitting media effects

K. Filintoglou[1], N. Papadopoulos[1], J. Arvanitidis[1,*], D. Christofilos[2], O. Frank[3,4],

M. Kalbac[3], J. Parthenios[4], G. Kalosakas[4,5], C. Galiotis[4,5], and K. Papagelis[4,5]

[1]Physics Department, Aristotle University of Thessaloniki, 54124 Thessaloniki, Greece

[2]Physics Division, School of Technology, Aristotle University of Thessaloniki, 54124

Thessaloniki, Greece

[3]J. Heyrovsky Institute of Physical Chemistry, v.v.i., Academy of Sciences of the Czech

Republic, Prague, 8, 18223 Czech Republic

[4]Institute of Chemical Engineering Sciences, Foundation of Research and Technology-Hellas

(FORTH/ICE-HT), Patras, 26504, Greece

[5]Materials Science Department, University of Patras, 26504 Patras, Greece FORTH/ICE-HT,

Stadiou Street, 26504 Rion Patras, Greece

The pressure evolution of the Raman spectrum of graphene grown by chemical vapour deposition on polycrystalline copper is investigated with the use of a polar and a non-polar pressure transmitting medium (PTM). The G and 2D Raman bands exhibit similar pressure slopes for both PTM irrespectively of any unintentional initial doping and/or strain of the samples. Our analysis suggests that any pressure-induced charge transfer effects are negligibly small to influence the pressure response of graphene. This is determined by the mechanical stress due to the pressure-induced substrate contraction and the transfer efficiency of the latter to the graphene layer, as well as the PTM-graphene interaction. For the non-polar PTM, a peculiar pressure behavior of graphene is observed in the PTM solidification regime, resembling that of free standing graphene.

## I. INTRODUCTION

The unique structure and fascinating properties of graphene provide a broad field for fundamental research and nanotechnology applications where low-cost, high quality, large-area graphene films are required. Nowadays, these specifications are met by the chemical vapor deposition (CVD) growth of graphene on polycrystalline metallic substrates like Cu or Ni.[1,2,3] Furthermore, strain engineering of graphene, allowing for the tailoring of its electronic properties by inducing different amounts and types of strain (uniaxial, biaxial or hydrostatic), is now poised to become a new exciting avenue of graphene research.[4] Raman spectroscopy, owing to its sensitivity on the structural and electronic characteristics of graphene, has been proven to be a valuable non-destructive in-situ tool in this emerging research area.[5,6,7] Indeed, the strain and the doping state of graphene have been evaluated by probing the changes of the so-called G and 2D characteristic Raman bands.[6,7,8,9,10,11,12]

Regarding the hydrostatic pressure Raman response of graphene, only two studies in mechanically exfoliated single- and few-layer graphene supported on $SiO_2/Si$ have appeared in the literature so far, employing different pressure transmitting media (PTM).[13,14] Proctor et al.[13] have additionally explored the pressure response of unsupported mixture of graphene flakes of different thickness. They concluded that single layer graphene (SLG) on $SiO_2/Si$ adheres well on the substrate and its compression at low pressures is determined by the compressibility of the substrate. On the other hand, Nicolle et al.[14] have obtained significantly smaller pressure slopes for the SLG Raman bands. In the case of the alcohol PTM, they have ascribed part of the pressure slope to considerable pressure mediated graphene doping from the substrate, owing to the polar nature of alcohol. Further experiments are needed to shed more light into the role of the graphene-substrate interaction in relation to the polarity of the PTM. In this context, we report here our results of high pressure Raman studies of CVD grown graphene on Cu substrate, using two different, polar and non-polar, PTM.

## II. EXPERIMENTAL AND CALCULATION DETAILS

The graphene sample was grown by CVD on a 25 μm thick polycrystalline copper foil (2x5 cm$^2$) in a quartz tube furnace system according to previously described procedure.[15] In brief, the copper foil was heated to 1000 °C and annealed for 20 min under H$_2$ gas flowing at 50 sccm (standard cubic centimeters per minute). The film was exposed to H$_2$ and CH$_4$ for 25 minutes, and 10 minutes to pure hydrogen. Finally, the substrate was cooled down under H$_2$.

Raman spectra were recorded in the backscattering geometry using a triple monochromator system (DILOR XY) equipped with a nitrogen-cooled charge coupled detector. The 514.5 nm line of an Ar$^+$ laser was used for excitation, focused on the sample by means of a x50 objective at a power lower than 2 mW to avoid any laser heating induced effects. The high pressure experiments were conducted in a Mao-Bell type diamond anvil cell (DAC) while the ruby fluorescence technique was used for pressure calibration. Two different mixtures, 4:1 methanol-ethanol (polar) and 1:1 FC70-FC77 fluorinert (non-polar), were used alternatively as PTM.

For the molecular dynamics simulations we have employed in-plane bond stretching and angle bending potentials derived from first principles. The stretching potential is a two-body Morse-type while the angle bending potential contains a typical quadratic term and an extra non-linear cubic one. [16] Rectangular graphene lattices have been simulated containing 7482 atoms while runs with larger lattice sizes provided identical results. In order to simulate hydrostatic pressure conditions, same magnitude forces per unit length are applied on each boundary atom. Consequently, forces of different magnitude are applied perpendicular to the zigzag and armchair edges.

## III. RESULTS AND DISCUSSION

Ambient pressure Raman spectra recorded from different locations of the graphene/Cu sample are illustrated in Figure 1 along with that of highly oriented pyrolytic graphite (HOPG) for comparison. The most prominent features in the Raman spectra of graphene are the so-called G and 2D bands. The G band at ~1588 cm$^{-1}$ originates from a first order Raman scattering process and it is associated with the doubly degenerate (iTO and LO) phonon mode (E$_{2g}$ symmetry) at the Brillouin zone center. On the other hand, the 2D band at ~2685 cm$^{-1}$ (for $\lambda_{exc}$= 514.5 nm) originates from a second order process, involving two iTO phonons near the K point of the Brillouin zone.[17] The frequencies, lineshapes and the relative intensity ratios I$_{2D}$/I$_G$ of the G and 2D bands in our samples along with the absence of the disorder-induced D band at ~1345 cm$^{-1}$ (for $\lambda_{exc}$= 514.5 nm), reveal that the vast majority of the sample comprises high quality single layer graphene.[1,2,5] Nevertheless, as inferred from Figure 1, the G and 2D band characteristics are position dependent. Namely, the obtained frequency, linewidth and I$_{2D}$/I$_G$ ratio variation ($\omega_G$= 1583-1594 cm$^{-1}$, $\Delta\omega_G$= 13-25 cm$^{-1}$, $\omega_{2D}$= 2677-2692 cm$^{-1}$, $\Delta\omega_{2D}$= 28-48 cm$^{-1}$, I$_{2D}$/I$_G$= 3-7.5) indicate non-uniform spatial doping and strain of the sample.[6,8,18].

The pressure response of the graphene/Cu samples is depicted in Figures 2 and 3. With increasing pressure, both Raman bands shift to higher frequencies without any significant changes in their width and relative intensity. For the fluorinert mixture, their frequency pressure dependence (squares in Fig. 3) is linear up to the pressure where abrupt changes occur (*vide infra*). The spatial inhomogeneity of the graphene/Cu sample is clearly reflected in the different 2D frequencies (~2680 and ~2689 cm$^{-1}$ at ambient pressure) for the two pressure runs performed for this PTM. Despite this difference, the pressure slopes for the two independent runs is practically the same for the 2D (21.3 and 22.0 cm$^{-1}$/GPa) and the G band (9.3 cm$^{-1}$/GPa). For the alcohol mixture, the frequency pressure dependence of both bands (circles in Fig. 3) appears sublinear for pressure up to 6 GPa that may be partially attributed to the stiffening of the graphene/Cu system at elevated pressures. In the low pressure regime up to 3 GPa (to be comparable to the fluorinert PTM data), the linear fitting of the data yields 9.1 $cm^{-1}$/GPa for the G and 18.3 $cm^{-1}$/GPa for the 2D band. Evidently, the pressure response of both bands is similar for both the PTM used (polar and non-polar). This finding along with the fact that the pressure slopes obtained in fluorinert are the same for samples with different initial doping, suggest that the observed pressure slopes of the Raman band frequencies are dominated by the mechanical stress and the initial or pressure induced doping effects (if any) have negligible contribution.

The mechanical and the doping related contributions can be distinguished by means of the $\omega_{2D}$-$\omega_G$ correlation plot following the analysis of Lee et al. for mechanically exfoliated graphene on $SiO_2$/Si at ambient conditions before and after thermal treatment.[11] In Figure 4, such a plot is presented using the frequencies of the G and the 2D bands obtained from our high pressure Raman spectra. According to Lee et al.,[11] a $\partial\omega_{2D}/\partial\omega_G$ slope of 2.2 is expected when only mechanical strain is considered, while values of $\omega_G$= 1581.6 and $\omega_{2D}$= 2676.9 $cm^{-1}$ were deduced for strain-free, undoped graphene (the star in Fig. 4). On the other hand, the carrier density effect on the frequencies of the G and the 2D bands can be obtained from the electrochemical n- and p-doping study of graphene by Das et al.,[7] assuming electron doping of CVD grown graphene on Cu (thick solid line in Fig. 4).[19,20,21,12] Pure n-doping for electron concentrations up to ~$2.5\times10^{13}$ $cm^{-2}$ would shift ($\omega_G$,$\omega_{2D}$) points almost horizontally in Figure 4. Note that possible p-doping of the sample due to its exposure in ambient air and moisture conditions[22,23,24] will not significantly alter this picture (thick dotted line in Fig. 4). As inferred from the figure, pressure shifts the ($\omega_G$,$\omega_{2D}$) data obtained for different doping/strain states at ambient pressure along constant electron density lines (lines of 2.2 slope). This is in accordance with our aforementioned statement that the pressure response of the Raman bands is dominated by the mechanical stress and furthermore clarifies that pressure does not cause any considerable doping variation of the sample irrespectively of the PTM. In addition, the $\omega_G$-$\omega_{2D}$ plot of Figure 4 allows the estimation of the doping state of the graphene samples in the DAC from the intersections of the constant electron density lines with the thick solid line representing pure n-doping. Hence, the electron density is ~0 and ~$3.5 \times 10^{12}$ cm$^{-1}$ for the two fluorinert PTM runs and ~$7.5 \times 10^{12}$ cm$^{-2}$ for the alcohol PTM.

Pressure data obtained from Nicolle et al.[14] for mechanically exfoliated SLG on SiO$_2$/Si substrate using the 4:1 methanol-ethanol mixture PTM are also included in Figure 3(a) for comparison. At ambient pressure, their $\omega_G$ downshift by ~8 cm$^{-1}$ with respect to our data should be attributed to the different doping/strain state of the samples. On the other hand, the significant redshift of the 2D band (by 40-50 cm$^{-1}$) in the SLG/SiO$_2$/Si case ($\lambda_{exc}$= 647.1 nm) with respect to our experiments ($\lambda_{exc}$= 514.5 nm) is attributable to the additional contribution of the excitation energy dispersive behavior of this band.[25] The linear fitting of the G band frequency for the SLG/SiO$_2$/Si sample in the low pressure regime (up to 3 GPa to allow comparison with our data) yields $\partial\omega_G/\partial P \approx 9$ cm$^{-1}$/GPa and $\partial\omega_{2D}/\partial P \approx 19$ cm$^{-1}$/GPa, very similar to those in our experiments, although Nicolle et al. concluded that their pressure coefficients are partly due to charge effects unlike our case.

In order to understand the observed pressure coefficient of the G band of graphene supported on various substrates, we recall that due to the monatomic thickness of the graphene layer, its compression is determined by the bulk modulus of the substrate as long as the graphene layer perfectly follows the pressure-induced substrate contraction.[13] Under this assumption and using a bulk modulus of 140 GPa for Cu[26] and a Grüneisen parameter $\gamma_{E2g}$ = 1.99 for the E$_{2g}$ mode (G band),[27] we expect a value of 15.0 cm$^{-1}$/GPa for $\partial\omega_G/\partial P$ (dotted line in Fig.3b). Similarly, according to Proctor et al.[13] the predicted pressure slope of the E$_{2g}$ mode is even larger for the more compressible SiO$_2$/Si substrate (~21.4 cm$^{-1}$/GPa), in agreement with their experimental findings in the low pressure regime (P<1.5 GPa). On the other hand, at higher pressures, they have attributed the lower pressure slope of the G band frequency to debonding and poorer adherence between graphene and $SiO_2$.[14] As pressure-induced charge transfer effects are negligible, the significant deviation of our experimental pressure slope from the calculated value of 15 $cm^{-1}$/GPa can only be ascribed also to non-ideal adherence of graphene on the Cu substrate, but from the beginning of the pressurization process. The non-ideal strain transfer from the more compressible substrate to the graphene layer leads to their relative sliding and thus to the reduced pressure slopes obtained. The similarity of the $\partial\omega_G/\partial P$ values in our experiments on Cu substrate and that of Nicolle et al.[14] on $SiO_2$/Si substrate (~9 $cm^{-1}$/GPa) with respect to the larger value anticipated for $SiO_2$/Si than that for Cu, suggests easier relative sliding in the case of the $SiO_2$/Si substrate. The relative sliding between two layers depends on the spatial gradient of the adhesion energy, which does not necessarily scale with the adhesion energy itself.[28,29] However, in the present case there is a scaling with the adhesion energy as its experimentally determined value for large-area monolayer graphene on Cu was as 0.72 ± 0.07 $Jm^{-2}$,[30] larger than 0.45 ± 0.02 $Jm^{-2}$ measured for single-layer graphene on $SiO_2$/Si.[31]

For the case of the fluorinert PTM, an abrupt decrease of the G band frequency is observed, accompanied by the reduction of the pressure slope (Fig. 3b). These changes take place for pressures higher than 2 and 2.5 GPa for the two independent pressure runs, respectively. They are irreversible and the data recorded upon pressure decrease follow the trend of the high pressure upstroke data, described by a reduced slope of 5.6 $cm^{-1}$/GPa. This intriguing pressure response in the case of the fluorinert PTM can be attributed to the interplay between the PTM- and the substrate-graphene interactions and its pressure evolution, both related to the specific chemical species involved. The picture becomes more complicated due to PTM solidification at higher pressures. In our case, the solidification pressure of the fluorinert mixture is ~1 GPa with quasi-hydrostatic behavior up to 2-3

GPa,[32,33] much lower than ~10.5 GPa where the glass transition of the alcohol mixture occurs.[34,35] When the PTM becomes solid, graphene finds itself in between two solid surfaces but still adhered to Cu. Upon further increase of pressure, one may assume that the interaction between graphene and the PTM surface tends to become comparable to that between graphene and Cu. At this point, and in combination with local non-hydrostatic components after the PTM solidification, graphene should not be considered as preferably adhered to copper anymore, resembling the pressure response of free-standing (unsupported) graphene. Indeed, our reduced pressure slope of 5.6 $cm^{-1}$/GPa coincides well with those obtained experimentally for unsupported mixture of graphene flakes of different thickness using nitrogen as PTM (~5 $cm^{-1}$/GPa) [13] or single-layer graphene using the alcohol PTM (5.6-5.9 $cm^{-1}$/GPa).[36] We note that the different critical pressure values where the supported-to-unsupported transition of graphene takes place in our experiments can be related to the spatial non-uniformity in the graphene-substrate interaction and in the strain.

Upon pressure reduction and for pressure lower than 2 GPa, a new peak appears in the Raman spectrum of graphene at a frequency slightly higher than that of the G band (Fig. 2b) with spatially dependent intensity. It can be attributed to the D′ defect related band of graphitic materials (~1620 $cm^{-1}$ at ambient pressure),[17] indicating the strong structural distortion of the graphene layer. Note, that the D band (~1345 $cm^{-1}$ at ambient pressure for $\lambda_{exc}$= 514.5 nm), fingerprint of disorder,[17] could not be recorded inside the DAC due to the particularly strong $T_{2g}$ first order Raman peak of diamond appearing at ~1332 $cm^{-1}$ at ambient pressure.[37] However, the D band is the dominant feature in the Raman spectrum of the recovered sample at ambient conditions after DAC opening (bottom spectrum in Fig. 2). In terms of the aforementioned discussion, this observation can be rationalized considering the comparable adherence of graphene to both copper and solid fluorinert at higher pressures. In this state, decompression results in graphene being preferentially adhered -depending on the local interaction and strain- to either copper or fluorinert, leading to its random ripping and significant structural distortion evidenced in the Raman spectrum.

Assuming our experimental value of 5.6 cm$^{-1}$/GPa as the slope of the G band for unsupported graphene, we can deduce the Grüneisen parameter from its definition for a quasi-harmonic vibrational mode, $\gamma_G = -\ln\omega_G/\ln V = B/\omega_G \cdot (d\omega_G/dP)$. This necessitates the knowledge of the bulk modulus, B of graphene. In this direction we have employed molecular dynamics simulations as described in the previous section. Figure 5 illustrates the force per unit length as a function of the relative surface change $\Delta S_R = (S-S_0)/S_0$, where $S_0$ is the initial undeformed surface of the system and S is the final equilibrium surface corresponding to the applied forces. Note the asymmetric response on hydrostatic tension (positive $\Delta S_R$ values) and compression (negative $\Delta S_R$ values) due to the profound asymmetry of the C-C interatomic stretching Morse-type potential. The positive end of the response curve corresponds to graphene fracture (~30% $\Delta S_R$),[16] whereas there is no fracture regarding the negative part of the curve. For sufficiently small stresses (the maximum stress applied in our experiments is less than 2 Nm$^{-1}$ or $\Delta S_R$ no more than -1%), the response is symmetric in compression and tension and the slope of the curve shown in the inset of Figure 5 provides a reliable value of graphene's 2D bulk modulus $B_{2D} \approx 200$ Nm$^{-1}$. It should be stressed that in order to describe properly the compression behavior of free standing graphene it is necessary to incorporate out-of-plane atomic displacements. However, the lateral support to graphene provided by the substrate and the PTM along with the perfect linear response of the Raman bands upon pressure application ensures that the mechanical response of graphene is solely determined by the in-plane potentials. Taking into account the thickness of a graphene monolayer ($l = 0.335$ nm, which is the interlayer spacing of graphite), the obtained effective 3D bulk modulus is $B_{eff} = B_{2D}/l \approx 600$ GPa. This value is in excellent agreement with that obtained from Monte

Carlo simulations.[38] Therefore, the Grüneisen parameter for graphene can be estimated as $\gamma_G$= 2.1, in agreement with biaxial and uniaxial experiments[39] and theoretical predictions.[27]

Finally, for the case of the alcohol PTM, the pressure increase should also lead to the decrease of the ratio of the PTM-graphene to the substrate-graphene interaction and possibly to the subsequent free-standing graphene behavior at sufficiently high pressure. In this framework, the relatively strong sublinear pressure response of the G band frequency observed in the alcohol PTM can be attributed to the enhancement of the interaction ratio at elevated pressures.

## IV. CONCLUSION

Summarizing, our high pressure Raman experiments of graphene on Cu suggest that there is no pressure-induced charge transfer for either polar or non-polar PTM. Furthermore, the initial strain/doping state of graphene does not affect the pressure response of the graphene bands. This is albeit determined by mechanical stress due to substrate contraction and thus the graphene adherence on the substrate and the compressibility of the latter. Of paramount importance is also the pressure dependent graphene-PTM interaction that may lead to a free-standing response of the graphene layer after sufficient pressurization. This behavior, in combination with molecular dynamics calculations of the graphene bulk modulus, allows the determination of the Grüneisen parameter of the $E_{2g}$ mode of graphene.

## ACKNOWLEDGEMENTS

The financial support by the Research Committee of the Aristotle University of Thessaloniki through the program "Support of Research Activities in A.U.Th." (grant No 89570) as well as that by the General Secretariat for Research and Technology of the Hellenic Ministry of Education and Religious Affairs through the program 'Deformation yield and failure of graphenes and graphene based nanocomposites' (ERC-10) are greatly acknowledged. M.K. acknowledges the support from the Czech grant agency (contract No.: P204/10/1677).

**Figure Captions**

**Figure 1.** Ambient pressure Raman spectra of CVD grown graphene on polycrystalline Cu substrate obtained from different sample positions along with that of highly oriented pyrolytic graphite (HOPG) for comparison. Dashed vertical lines denote the lowest frequencies of the G and the 2D band presented in the graph.

**Figure 2.** Raman spectra of CVD grown graphene on Cu substrate at various pressures using the 1:1 FC70-FC77 fluorinert mixture as PTM. (a) Spectra recorded upon pressure increase. Dashed vertical lines denote the initial frequencies of the G and the 2D bands. (b) Spectra recorded upon pressure decrease after subtraction of the luminescent background originating from the Cu substrate. Asterisks mark peaks of the diamond anvil. The bottom spectrum was recorded after complete pressure release and the subsequent DAC opening.

**Figure 3.** (a) The frequencies of the G and 2D bands of CVD grown graphene on Cu substrate as a function of pressure for the two PTM used (alcohol and fluorinert mixture) in comparison with data obtained from the literature for single-layer graphene on $SiO_2$/Si substrate using the 4:1 methanol-ethanol mixture PTM (asterisks in the figure).[14] (b) Pressure dependence of the G band frequency. Open (solid) symbols correspond to data obtained upon pressure increase (decrease). The dotted line refers to the theoretically expected pressure dependence of the G band of graphene assuming that perfectly follows the pressure-induced contraction of Cu substrate. Numbers in both panels refer to the pressure slopes of the Raman bands up to ~3 GPa, while lines through the experimental data are least square fittings.

**Figure 4.** Correlation between the frequencies -as obtained from the high pressure Raman spectra- of the G and 2D bands of graphene on Cu substrate. Open and solid squares correspond to the two pressure runs employing the fluorinert PTM up to 3

GPa, while circles to alcohol PTM up to 6 GPa. Dashed lines through the experimental data are constant carrier density (in cm$^{-2}$) lines, while the star symbol refers to undoped, strain-free graphene, following Lee et al.[11] The thick solid (dotted) line, appropriately shifted to start from the undoped/strain-free point,[11] represents $\omega_{2D}$-$\omega_G$ correlation for electrochemically electron (hole) doped graphene with data obtained from Das et al.[7]

**Figure 5.** Force per unit length as a function of the relative surface change $\Delta S_R = (S - S_0)/S_0$ as deduced from molecular simulations for graphene.

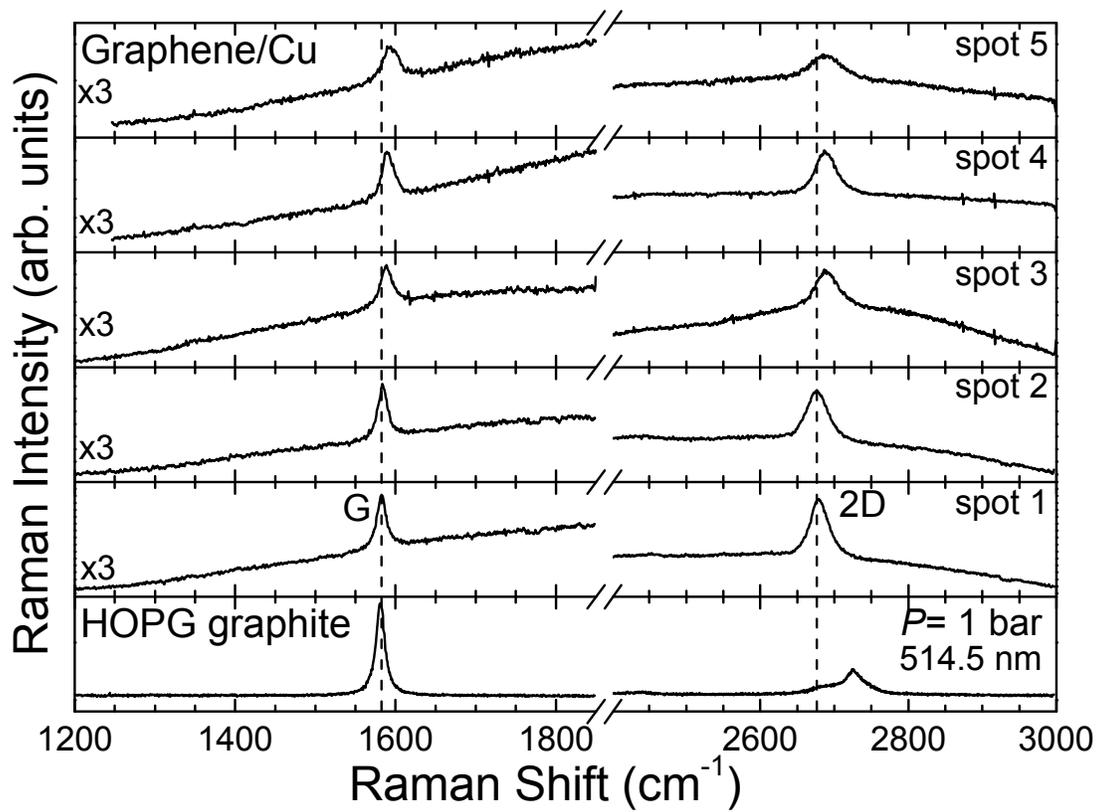

Fig. 1

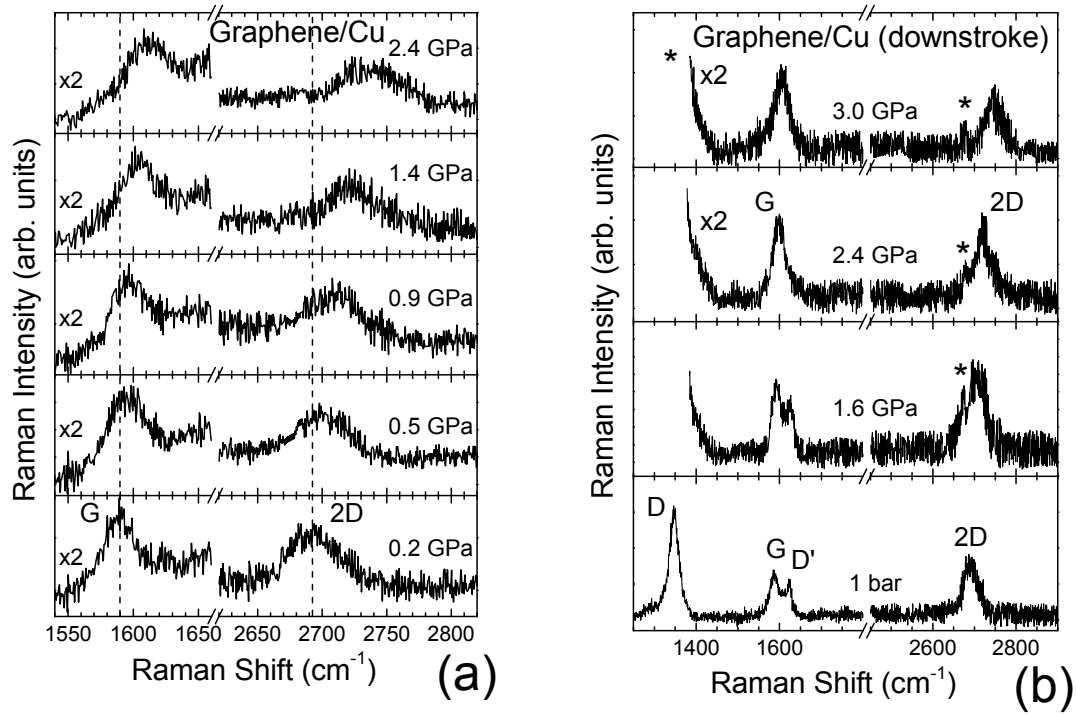

**Fig. 2**

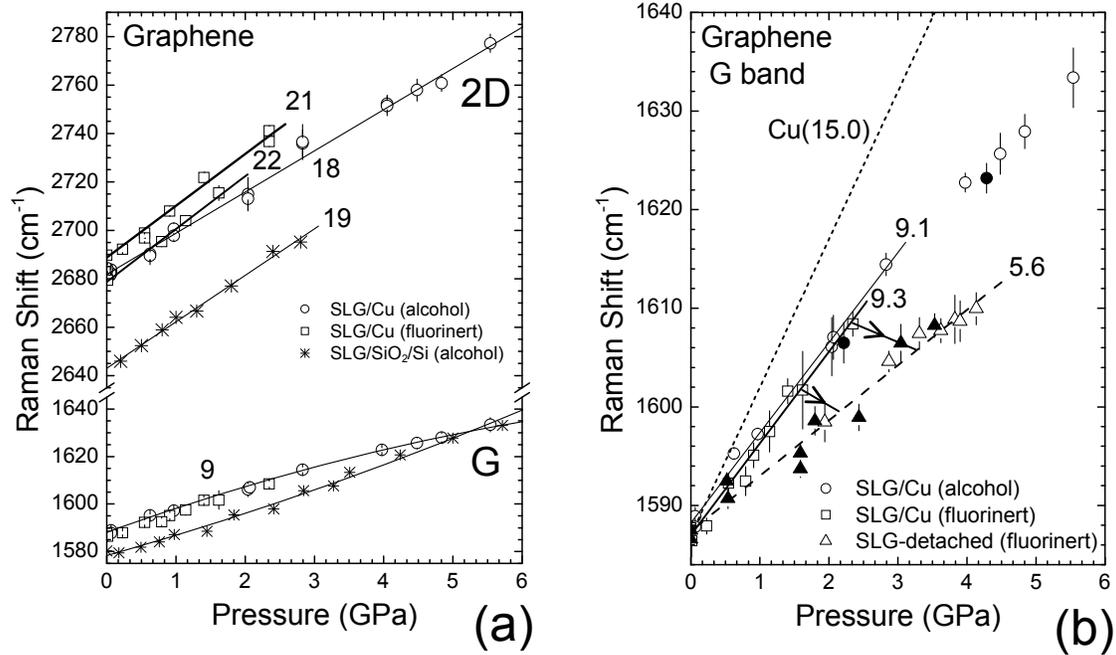

**Fig. 3**

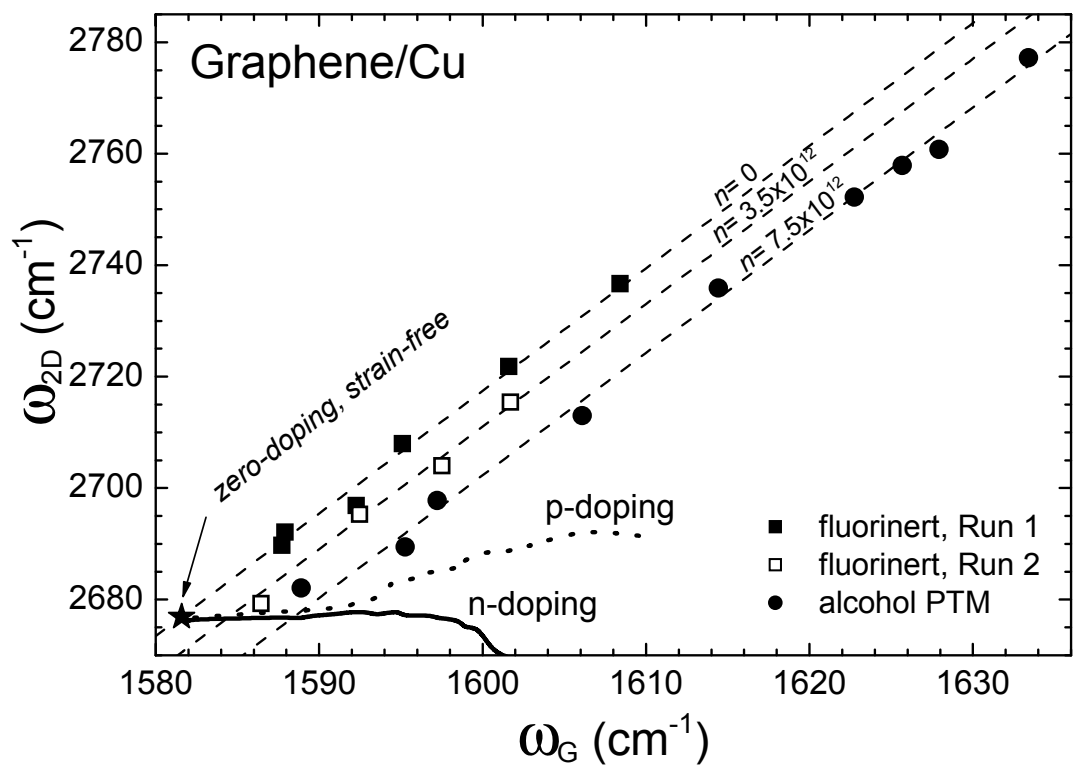

Fig. 4

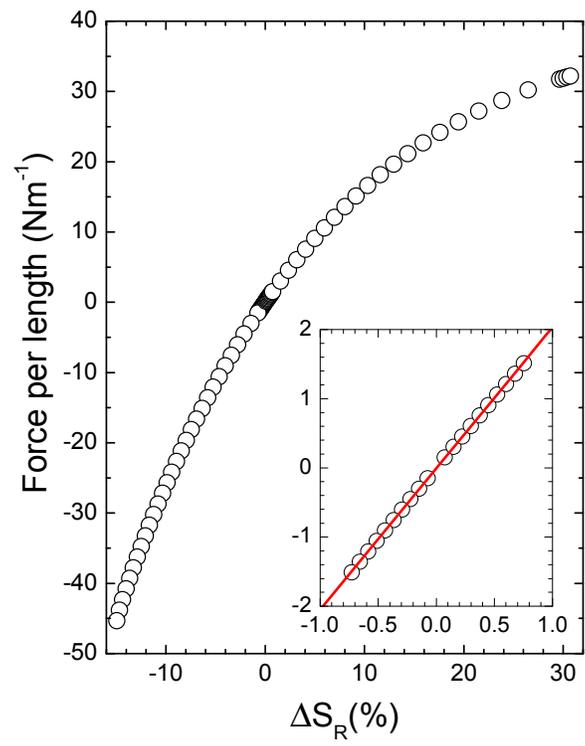

**Fig. 5**